\documentclass[11pt]{article}
\usepackage{amsmath,amssymb,color,epsfig,cite}
\usepackage{graphicx}
\usepackage{subfigure}
\usepackage{setspace}

\usepackage{amsfonts}

\newcommand{\hoch}[1]{$\, ^{#1}$}

\makeatletter
\@addtoreset{equation}{section}
\makeatother

\newcommand{\be}{\begin{equation}}
\newcommand{\ee}{\end{equation}}
\newcommand{\bea}{\setlength\arraycolsep{2pt} \begin{eqnarray}}
\newcommand{\eea}{\end{eqnarray}}
\newcommand{\nn}{\nonumber}

\def\ft#1#2{{\textstyle{\frac{\scriptstyle #1}{\scriptstyle #2} } }}
\def\fft#1#2{{\frac{#1}{#2}}}

\def\0{{\sst{(0)}}}
\def\1{{\sst{(1)}}}
\def\2{{\sst{(2)}}}
\def\3{{\sst{(3)}}}
\def\4{{\sst{(4)}}}
\def\5{{\sst{(5)}}}
\def\6{{\sst{(6)}}}
\def\7{{\sst{(7)}}}
\def\8{{\sst{(8)}}}
\def\9{{\sst{(9)}}}

\def\sst#1{{\scriptscriptstyle #1}}

\def\del{{\partial}}

\begin{document}



\begin{center}
{\large {\bf An $a$-theorem for Horndeski Gravity at the Critical Point}}

\vspace{10pt}
Yue-Zhou Li\hoch{\dag} and  H. L\"u\hoch{*}

\vspace{15pt}

{\it Department of Physics, Tianjin University, Tianjin 300072, China}

\vspace{30pt}

\underline{ABSTRACT}
\end{center}

We study holographic conformal anomalies and the corresponding $a$-theorem for Einstein gravity extended with Horndeski terms that involve up to and including linear curvature tensors.  We focus on our discussion in $D=5$ bulk dimensions. For the generic Horndeski coupling, the $a$-charge is the same as that in Einstein gravity, but the inclusion of the Horndeski term violates the $a$-theorem.  However, there exists a critical point of the Horndeski coupling, for which the theory admits nearly AdS spacetimes with non-vanishing Horndeski scalar. The full AdS isometry is broken down by the logarithmic scalar hair to the Poincar\'e group plus the scale invariance.  We find that in this case the $a$-charge depends on the AdS radius $\ell$ and the integration constant $\chi_s$ of the Horndeski scalar.  In addition, we find that two new central charges emerge, that are absent in gravities with minimally-coupled matter. We call them $b$-charges. These $b$-charges also depend on $\ell$ and $\chi_s$.  We construct an $a$-function for fixed $\ell$ but with the running Horndeski scalar $\chi$ replacing the constant $\chi_s$, and establish the holographic $a$-theorem using the null energy condition in the bulk.  Furthermore, we find that there exist analogous  monotonous $b$-functions as well.  We also obtain the $a$-charge and the $a$-theorem in general odd bulk dimensions.

\vfill {\footnotesize \hoch{\dag}liyuezhou@tju.edu.cn\ \ \ \hoch{*}mrhonglu@gmail.com\ \ \ }

\pagebreak

\tableofcontents
\addtocontents{toc}{\protect\setcounter{tocdepth}{2}}


\newpage

\section{Introduction}

The AdS/CFT correspondence \cite{Maldacena:1997re} serves as a powerful tool to investigate a certain strongly coupled conformal field theory (CFT) in $d$ dimensions by studying its weakly coupled dual classical anti-de Sitter (AdS) gravity in $D=d+1$ dimensions.\footnote{In this paper, we shall always use $D$ to refer the dimensions of bulk gravity, and $d$ to refer the dimensions of boundary field theory.} In particular, one can employ the gauge/gravity duality to determine the conformal anomaly of the CFT by means of the Fefferman-Graham (FG) expansion in bulk gravity in $D=2n+1$ dimensions \cite{Henningson:1998gx,Henningson:1998ey,Imbimbo:1999bj,Schwimmer:2003eq}. The resulting holographic conformal anomaly may arise as the trace of the boundary energy momentum tensor that would be non-vanishing even on the vacuum. It turns out that the holographic conformal anomaly in the gravitational sector is generally given by \cite{Imbimbo:1999bj}
\be\label{Weyl}
\langle T_\mu{}^\mu{}\rangle\sim -a E^{(2n)}+\sum_{i}c_{i}I_{i}^{(2n)}\,,
\ee
where $E^{(2n)}$ is the Euler density and $I_{i}^{(2n)}$'s are all the Weyl invariants in $d=2n$ dimensions. Eq.~(\ref{Weyl}) is the most general formula for the holographic conformal anomaly arising from a pure gravity action.  Any gravity in odd dimensions constructed from arbitrary higher-order curvature invariants exhibits the holographic conformal anomaly as in  (\ref{Weyl}). The details of the theory are encoded in the specific constant coefficients $a$ and $c_i$, which represent two different types of central charges. These charges are typically expressed in terms of various coupling constants of higher-order terms \cite{Nojiri:1999mh,Blau:1999vz,Banados:2004zt,Kraus:2005zm,
Banados:2005rz,Banerjee:2009fm,Myers:2010ru,Myers:2010jv,Myers:2010xs,Myers:2010tj,
Li:2017ncu,Li:2017txk}. The form of the holographic conformal anomaly (\ref{Weyl}) coincides precisely with that obtained from the explicit CFT calculation \cite{Duff:1977ay,Duff:1993wm,Bonora:1985cq}.

Wilsonian description of quantum field theory (QFT) states that higher energy modes are integrated out along the renormalization group (RG) flow from higher energy to lower energy and the degrees of freedom decrease irreversibly. The quantitative statement is typically referred as the $a$-theorem, namely, there exist $a$-charges measuring the massless degrees of freedom of the CFT at the RG fixed points, and the charge at the ultra-violet (UV) fixed point is always larger than or equal to that at the infra-red (IR) fixed point:
\be
a_{\rm UV}\geq a_{\rm IR}\,.
\ee
An even stronger statement is that there exists an $a$-function of energy scale $\mu$ that becomes $a_{\rm UV}$ and $a_{\rm IR}$ at the respective fixed points, and the function $a(\mu)$ is monotonically increasing, namely
\be
\fft{d}{d\mu}a(\mu)\geq0\,.
\ee
The first important example for $a$-theorem is the $c$-theorem of $d=2$ CFT established by Zamolodchikov \cite{zamo}. Further works in higher dimension can be found in \cite{Cardy:1988cwa,Komargodski:2011vj,Huang:2015sla}.

The $a$-theorem can also be investigated in the context of AdS/CFT correspondence \cite{Girardello:1998pd,Freedman:1999gp,Myers:2010xs}, in which the energy scale corresponds to bulk radius $r$
\be
\fft{d}{dr}a(r) \ge 0\,,\qquad a(r)\Big|_{\rm AdS}=a\,,\label{atheorem}
\ee
for certain $a(r)$ function, where $r\rightarrow\infty$ is the boundary of the asymptotic AdS geometry, which corresponds to the UV region of the dual CFT. The constant $a$ above is the coefficient in front of the Euler density evaluated at the AdS fixed point. In the holographic framework, one could readily generalize the holographic $a$-theorem to arbitrary dimensions \cite{Myers:2010tj,Li:2017txk}, and for gravity theory involving higher-order curvature invariants, the general dimension version of the holographic $a$-theorem, in turn, constrains the coupling constants \cite{Li:2017txk}. More properties of RG flow other than $a$-theorem can also be studied in the framework of holography, say \cite{deBoer:2000cz,Skenderis:2002wp,Rajagopal:2015lpa}.

Matter fields may also contribute to the holographic conformal anomalies. The contribution from minimally coupled scalar fields were studied in \cite{Nojiri:1998dh,Nojiri:2000kh,Papadimitriou:2011qb}. However, the new anomalies do not alter either the form or explicitly $a$ and $c$ charges in (\ref{Weyl}) in the gravitational sector, but rather they involves boundary matter fields, and we call them anomalies in the matter sector.  The situation becomes different when we consider non-minimally coupled matter and novel holographic feature may arise. This is because non-minimal couplings mix gravity and matter, which may alter the asymptotic behavior of the matter fields, and hence the FG expansions, yielding novel contributions to the holographic anomaly.

In this paper, we consider Einstein-Horndeski gravity.  Horndeski invariants were constructed from curvature tensors and axionic scalars \cite{Horndeski:1974wa}.  They were later rediscovered \cite{Nicolis:2008in} as some Galilean theories, and were deeply investigated in its application in cosmology, see e.g.~\cite{Amendola:1993uh,Germani:2010gm}.  For our purpose, we
focus on Einstein gravity with a negative cosmological constant, together with the kinetic term for the axion $g_{\mu\nu}\partial^{\mu}\chi\partial^{\mu}\chi$ and one Horndeski term $G_{\mu\nu}\partial^{\mu}\partial^{\nu}\chi$. The vacuum is the AdS spacetime where the scalar vanishes.
It turns out that there exist a critical point of the Horndeski coupling, for which a nearly AdS spacetime can emerge where the scalar is non-vanishing. Properties and applications on beyond cosmology has also been actively pursued. Locally asymptotically AdS black hole solutions were constructed \cite{Rinaldi:2012vy,Anabalon:2013oea,Babichev:2013cya,Cisterna:2014nua,Jiang:2017imk,Feng:2017jub}; Stability and causality were discussed in \cite{Jimenez:2013qsa,Kobayashi:2014wsa,Minamitsuji:2015nca}; black hole thermodynamics were analysed in \cite{Feng:2015oea,Feng:2015wvb}; the AdS/CMT properties were given in \cite{Jiang:2017imk,Baggioli:2017ojd,Liu:2017kml}; further properties and applications were studied in \cite{Caceres:2017lbr,Feng:2017jub,Geng:2017nwv}.

In this paper we examine the holographic $a$-theorem for the Einstein-Horndeski gravity. In $D=5$ dimensions, we derive the holographic conformal anomaly using the FG expansions and find that result is ${\cal A}_{\rm gr}+ {\cal A}_{m}$, The quantity ${\cal A}_m$ is given in terms of $\chi^{(0)}$, the boundary field of the Horndeski axion $\chi$. We thus refer ${\cal A}_m$ as the anomaly in the matter sector.  The anomalous term ${\cal A}_{\rm gr}$ is expressed in terms of the boundary curvature invariants and hence we refer it as the anomaly in the gravitational sector.  Intriguingly there is no mix between the two sectors, as if the scalar is minimally coupled, even though the Horndeski scalar is definitely not minimally coupled in the bulk.

For generic couplings of the Einstein-Horndeski gravity we consider in this paper, we find that the anomalous term ${\cal A}_{\rm gr}$ is identical to that of Einstein gravity.  In other words, the Horndeski scalar gives contributions only to ${\cal A}_m$, but no contribution to ${\cal A}_{\rm gr}$.
Furthermore, we find that there does not exist an $a$-function that can lead to the $a$-theorem (\ref{atheorem}).

However, there exists a critical point of the Horndeski coupling for which the theory admits a nearly AdS spacetime where the Horndeski scalar is non-vanishing.  The FG expansion for the scalar $\chi$ at the critical point admits an additional logarithmic mode.  This mode can give non-trivial contribution to the $a$ and $c$ charges.  Furthermore, we find that the structure of the anomalous term ${\cal A}_{\rm gr}$ in $D=2n+1$ in critical Einstein-Horndeski gravity is augmented from (\ref{Weyl}) to become
\be\label{Weyl2}
{\cal A}_{\rm gr} \sim -a E^{(2n)}+\sum_{i}c_{i}I_{i}^{(2n)} + \sum_k b_k H_{k}^{(2n)}\,.
\ee
In other words, new central charges emerge, and we shall call them $b$-charges. In $D=5$ dimensions, we perform explicit calculations and find that there are two $H^{(4)}$, given by
\be
H_1^{(4)} = \Box R\,,\qquad H_2^{(4)} = R^{ij} R_{ij} - \ft14 R^2 - \ft14 \Box R\,.
\ee
Thus in five dimensions, the inclusion of the Horndeski term at the critical coupling gives rise to a total of four holographic conformal charges in the gravitational sector, associated with all the four possible boundary curvature terms at the fourth-order. With appropriate recombination, they can be grouped as
\bea
{\cal A}_{\rm gr} &\sim& -a E^{(4)} + c I^{(4)} + b_1 H_1^{(4)} + b_2 H_2^{(4)}\,,\cr
E^{(4)} &=& R^2 - 4 R^{ij} R_{ij} + R^{ijkl} R_{ijkl}\,,\cr
I^{(4)} &=& \ft13 R^2-2R_{ij}R^{ij}+R_{ijkl}R^{ijkl}\,.\label{D5acbb}
\eea
By contrast, for gravity with only minimally coupled matter, there are only $a$ and $c$ charges, whilst the new $b$-charges are absent. Another distinguishing feature of the central charges for Einstein-Horndeski gravity at the critical point is that they depend not only on the coupling constants of the theory, but also the integration constant $\chi_s$ associated with the scalar hair.  In conformal field theory, the anomalous trace term of the form $\Box R$, i.e.~$H_1^{(4)}$ with constant coefficient can be removed by some appropriate local counterterm proportional to $R^2$. Interestingly, in the case of holographic conformal anomaly for Einstein gravity (or higher-order curvature gravity) with minimally-coupled matter, such a term does not even arise.  However, as we shall discuss in the next section, the appearance of the holographic anomalous term $H_1^{(4)}$ in Einstein-Horndeski gravity is not removable by a local counterterm in the boundary field theory owing to the fact that the charge depends on the scalar hair $\chi_s$.

The paper is organized as follows.  In section 2, we briefly review the Einstein-Horndeski theory.  We perform the FG expansion explicitly in $D=5$ and derive the conformal anomalies for both generic and critical couplings. In section 3, we establish that there is no $a$-theorem for the case with generic couplings.  For the critical theory, find an $a$-function and establish the $a$-theorem using the null energy condition.  In section 4, we study the conformal anomaly in general $D=2n+1$ dimensions.  Owing to the complexity of the calculations, we deduce only the $a$-charge using the technique developed in \cite{Li:2017txk}. We then establish an $a$-theorem using the null energy condition.  We conclude the paper in section 5.

\section{Holographic conformal anomaly in $D=5$}
\label{sec:D=5}

\subsection{Horndeski gravity and the FG expansions in general dimensions}
\label{sec:D=5Hornd}

We begin with reviewing Einstein-Horndeski gravity involving up to only linear curvature terms. The action in general dimensions $D$ is given by
\be
S=\fft{1}{16\pi}\int d^{D}x\sqrt{-g}\mathcal{L}\,,\qquad \mathcal{L}=R-2\Lambda_0-\ft{1}{2}(\alpha g_{\mu\nu}-\gamma G_{\mu\nu})
\del^\mu\chi\, \del^\nu\chi\,,\label{action}
\ee
where the bare cosmological constant $\Lambda_0$ is taken to be negative in general. The parameters $\alpha$ and $\gamma$ are coupling constants, and $G_{\mu\nu}=R_{\mu\nu}-\fft{1}{2}Rg_{\mu\nu}$ is Einstein tensor. Since $\chi$ appears in the action only through a derivative, there is a constant shift symmetry associated with $\chi$, implying that $\chi$ is axionic. Note that it is reasonable to expect that $\alpha >0$, in which case, we can set $\alpha=+1$ without loss of generality.  In this paper, we shall let $\alpha$ be arbitrary.

The equations of motions associated with the variations of the metric $g^{\mu\nu}$ and $\chi$ are respectively given by
\bea
E_{\mu\nu} &=& G_{\mu\nu} +\Lambda_0 g_{\mu\nu}-
\ft12\alpha \Big(\partial_\mu \chi \partial_\nu \chi - \ft12 g_{\mu\nu} (\partial\chi)^2\Big)-\ft12\gamma \Big(\ft12\partial_\mu\chi \partial_\nu \chi R - 2\partial_\rho
\chi\, \partial_{(\mu}\chi\, R_{\nu)}{}^\rho \cr
&&- \partial_\rho\chi\partial_\sigma\chi\, R_{\mu}{}^\rho{}_\nu{}^\sigma -
(\nabla_\mu\nabla^\rho\chi)(\nabla_\nu\nabla_\rho\chi)+(\nabla_\mu\nabla_\nu\chi)
\Box\chi + \ft12 G_{\mu\nu} (\partial\chi)^2\cr
&&-g_{\mu\nu}\big[-\ft12(\nabla^\rho\nabla^\sigma\chi)
(\nabla_\rho\nabla_\sigma\chi) + \ft12(\Box\chi)^2 -
  \partial_\rho\chi\partial_\sigma\chi\,R^{\rho\sigma}\big]\Big)=0\,,\cr
E_{\chi} &=&\nabla^\mu \big( (\alpha g_{\mu\nu} - \gamma G_{\mu\nu}) \nabla^\nu\chi\big)=0\,.
\label{EOM}
\eea
For vanishing $\chi$, these equations admit maximally symmetric vacuum AdS vacuum with $G_{\mu\nu}=-\Lambda_0 g_{\mu\nu}$. As we shall see later, at certain critical point of the
Horndeski coupling $\gamma$, there exists a nearly AdS spacetime for which $\chi$ is
non-vanishing and its integration constant contributes to the effective cosmological constant $\Lambda_{\rm eff}$.  We shall use $\Lambda_{\rm eff}$ for the general case and it becomes bare $\Lambda_0$
when $\chi=0$.  We now parameterize $\Lambda_{\rm eff}$ by
\be
\Lambda_{\rm eff}=-\fft{d(d-1)}{2\ell^2}\,,\label{lambdaeff}
\ee
where $\ell$ is the AdS radius. We emphasize again here that we shall always use $D$ to denote the total bulk dimensions and $d$ to denote the boundary dimensions.  In this vacuum, the effective kinetic term for Horndeski scalar $\chi$ is
\be
\mathcal{L}_{\chi}=-\ft{1}{2}(\alpha+\gamma\Lambda_{\rm eff})(\partial\chi)^2\,.
\ee
The absence of ghost excitation requires that $\alpha+\gamma\Lambda_{\rm eff}\ge 0$, with the equality corresponding to the critical point.  It follows from (\ref{lambdaeff}) that the critical point of the Horndeski coupling is given by \cite{Jiang:2017imk,Feng:2017jub}
\be
\gamma = \fft{2\ell^2}{d(d-1)}\alpha\,.
\label{critical}
\ee
Analogous critical points were also found in Einstein-Gauss-Bonnet gravity \cite{Fan:2016zfs}.

In order to investigate holographic RG flow in Einstein-Horndeski gravity, we analyze the asymptotic behaviour using the FG expansions.  For the metric ansatz of the form
\be
ds^2=\fft{\ell^2}{4\rho^2}d\rho^2+\fft{g_{ij}}{\rho}dx^i dx^j\,,\qquad i,j=1,\cdots, d\,,
\ee
the FG expansion for $g_{ij}$ is
\begin{equation}
\begin{split}
&g_{ij}=g^{\0}_{ij}+\rho g^{\2}_{ij}+\rho^{2}g^{\4}_{ij} + \cdots
\\&g^{ij}=g^{\0 ij}-\rho g^{\2 ij}-\rho^{2}\big(g^{\4 ij}-\ft{1}{4}{\rm Tr}(g^{\2 2})g^{\0 ij}\big)+ \cdots\,.
\end{split}
\label{FG}
\end{equation}
In this expansion, the $\rho=0$ represents the boundary of the AdS. Since the Einstein-Horndeski theory we study in this paper involves only Ricci tensors, we do not need to compute the FG expansions for the Riemann tensors.  The relevant components for Ricci tensors are
\bea
R_{ij}&=&\hat{R}_{ij}-\dfrac{2\rho}{\ell^2}g''_{ij}+\dfrac{1}{\ell^2}g^{lk}g'_{lk}g_{ij}+
\dfrac{2\rho}{\ell^2}g^{kl}g'_{lj}g'_{ki}
-\dfrac{\rho}{\ell^2}g^{lk}g'_{lk}g'_{ij}+\dfrac{d-2}{\ell^2}g'_{ij}-\dfrac{d}{\ell^2}
\dfrac{1}{\rho}g_{ij}\,,\nn\\
R_{\rho\rho}&=&-\dfrac{d}{4\rho^{2}}-\ft{1}{2}g^{ij}g''_{ij}+
\ft{1}{4}g^{ik}g^{jl}g'_{ij}g'_{kl}\,,
\eea
where a prime here denotes a derivative with respect to $\rho$ and $\hat R_{ij}$ is the Ricci tensor of the boundary metric $g_{ij}$.  Expanding $\hat R_{ij}$ in terms of $\rho$, we have
\be
\hat{R}_{ij}=\hat{R}^{(0)}_{ij}+\rho \hat{R}^{(2)}_{ij}+\cdots\,,
\ee
where $\hat{R}^{(2)}_{ij}$ is given in terms of $g^{(2)}_{ij}$
\be
\hat{R}^{(2)}_{ij}=-\ft12\nabla^{(0)}_{i}\nabla^{(0)}_{j} (g^{(0)kl}g^{(2)}_{kl})
-\ft12\nabla^{(0)}_{k}\nabla^{(0)k}g^{(2)}_{ij}
+g^{(0)kl}\nabla^{(0)}_{k}\nabla^{(0)}_{(i}g^{(2)}_{j)l}\,.\label{hatR2ij}
\ee
Its trace by $g^{(0)}_{ij}$ is
\be
\hat{R}^{(2)}=-\Box^{(0)}{\rm Tr}g^{(2)}+\nabla^{(0)i}\nabla^{(0)j}g^{(2)}_{ij}\,,\label{R2}
\ee
and hence $\sqrt{g^{(0)}} \hat{R}^{(2)}$ is a total derivative. Here ${\rm Tr} g^{(2)}=g^{(0)kl}g^{(2)}_{kl}$.

For asymptotically AdS spacetimes, the FG expansion for the Horndeski scalar $\chi$ takes the form
\be
\chi=\chi^{\0}+\rho\chi^{\2}+\rho^2 \chi^{\4}+\cdots\,.
\label{normalchi}
\ee
In the usual context of the AdS/CFT correspondence, this massless scalar is dual some scalar operator with conformal dimension $\Delta=4$.  To be specific, $\chi^\0$ plays the role of the source of the corresponding scalar operator, and its response appears starting at $\rho^2$ order.
As we shall see in the next subsection, the above expansion is consistent with the equations of motion for general coupling constants.  However, at the critical point (\ref{critical}), there exist asymptotically nearly AdS spacetimes where there is a new logarithmic term for $\chi$.  Thus at the critical coupling, the FG expansion for the scalar becomes
\be
\chi=\chi_{s}\log\rho+\chi^{\0}+\rho\chi^{\2}+\rho^2 \chi^{\4}+\cdots\,,
\label{criticalchi}
\ee
where $\chi_s$ is set to be a constant. Thus at the critical point, the additional scalar hair $\chi_s$ appears.  Although the metric is asymptotically AdS, the isometry $SO(2,4)$ is broken by this mode, and we call the asymptotic geometry nearly AdS, whilst the boundary field theory as nearly conformal. To be specific, the surviving symmetry is the Poincar\'e group together with a scaling invariance, which is a subgroup of $SO(2,4)$.  The corresponding dual theory becomes the scale-invariant but not conformally-invariant quantum field theory. This subject has been reviewed in \cite{Nakayama:2013is}. As we shall see in subsection \ref{sec:D=5crit}, this new logarithmic term in the critical theory gives nontrivial contribution to the holographic conformal anomaly.  In what follows, we shall focus on $D=5$ bulk dimensions, corresponding to $d=4$ boundary dimensions.

\subsection{Holographic conformal anomaly for generic couplings}
\label{sec:D=5generic}

In this subsection, we consider FG expansions (\ref{FG}) for $g_{ij}$ and (\ref{normalchi}) for $\chi$ respectively. Substituting all these into (\ref{action}) and take $D=5$, we end up with
\be
S=\fft{1}{16\pi}\int d^{4}x\int_{\epsilon} d\rho\sqrt{-g}\mathcal{L}=\fft{1}{16\pi}\int d^{4}x \sqrt{g^{(0)}} \int_{\epsilon} d\rho(\cdots+\fft{\mathcal{A}}{\rho}+\cdots)\,,\label{FGaction}
\ee
where the action has been expanded in Laurent series around $\rho=0$, the dots in the left and right of $\mathcal{A}/\rho$ represent negative and positive (and zero) powers of $\rho$ respectively, and the coefficient $\mathcal{A}$ will be given soon. The nearly AdS boundary is located at $\rho=\epsilon\sim0$ in the FG expansion. It is therefore obvious that the right-dot terms are convergent at the AdS boundary.  The left-dot terms are divergent, but can be removed by the appropriate Gibbons-Hawking type surface terms and the holographic counterterms.  The $\mathcal{A}/\rho$ term can only appear in {\it odd} bulk dimensions, and it gives rise to a logarithmic divergence at the AdS boundary after integrating over $\rho$. This divergence cannot be cancelled by either the Gibbons-Hawking or holographic counterterms since these surface terms have only the power-law divergence owing to the specific ansatz of the FG expansions.  This term is then interpreted as the holographic conformal anomaly.

After some involved algebra, we find that the anomalous term $\mathcal{A}$ is given by
\begin{eqnarray}
\ell^{-1}\mathcal{A}&=& \beta_1 {\rm Tr} g^{(4)}+A_1 {\rm Tr} (g^{(2)2})+A_2 ({\rm Tr} g^{(2)})^{2}-\ft12 ( \hat{R}^{(0)}_{ij}g^{(2)ij}-\hat{R}^{(2)})\cr
 &&+{\rm Tr} (g^{(2)})\big(\ft14\hat{R}^{(0)}+A_3 (\nabla^{(0)}\chi^{(0)})^2\big) + A_4 g^{(2)ij}\nabla^{(0)}_i \chi^{(0)}\nabla^{(0)}_j \chi^{(0)}\cr
 &&-\ell^{-2}\beta_2 \chi^{(2)2}
 -\ft{1}{2}\beta_2\, g^{(0)ij}\nabla^{(0)}_i \chi^{(2)}\nabla^{(0)}_j \chi^{(0)} +\ft{1}{4}\gamma \hat{G}^{(0)ij}\nabla^{(0)}_i \chi^{(0)}\nabla^{(0)}_j \chi^{(0)}\,.\label{actionA}
\end{eqnarray}
Various coefficients above are given by
\bea
&&\beta_{1}=-\ft{3}{\ell^2}-\ft12{\Lambda_0}\,,\qquad \beta_{2}=\alpha-\ft{6\gamma}{\ell^2}\,,\qquad
A_1=\ft{1}{\ell^2}+\ft14{\Lambda_0}\,,\nn\\
&&A_2=-\ft{1}{4\ell^2}-\ft18{\Lambda_0}\,,\qquad
A_3=-\ft18{\alpha}+\ft{\gamma}{4\ell^2}\,,\qquad A_4=\ft14{\alpha}-\ft{\gamma}{\ell^2}\,.
\label{D5coef1}
\eea
Note that $\sqrt{g^{(0)}} \hat{R}^{(2)}$ is a total derivative and hence does not contribute to the equations of motion, but it may contribute to the total anomaly.

     The equations of motion of the boundary fields in the FG expansions can be obtained by subsituting
the ansatz into the full equations of motion (\ref{EOM}) and solve them order by order in the powers of $\rho$.  This can be cumbersome.  Since the FG ansatz is self consistent, and we can treat
\be
S_{\rm eff} = \int d^4 x \sqrt{g^{(0)}} {\cal A}\,,
\ee
as an effective action and derive the equations of motion of the boundary fields.  We shall adopt this much simpler approach.

Equations of motion associated with the variation of $g^{(4)}_{ij}$ is $\beta_1 =0$, which relates the bare cosmological constant with the AdS radius, namely
\be
\Lambda_0=-\fft{6}{\ell^2}\,.
\ee
Thus in this case, we have $\Lambda_{\rm eff}= \Lambda_0$. Equations of motion associated with the variation of $\chi^{(2)}$ is then
\be
\fft{\beta_2}{\ell^2} \big(\chi^{(2)}-\ft14{\ell^2}\Box^{(0)}\chi^{(0)}\big)=0\,.
\ee
One way to solve the above equation is to set $\beta_2=0$, corresponding precisely to the critical condition (\ref{critical}) in $d=4$.  In this subsection, we shall consider generic couplings with $\beta_2\ne 0$, for which case the above equation is solved with
\be
\chi^{(2)}=\ft{1}{4}\ell^2\,\Box^{(0)}\chi^{(0)}\,.
\ee
The variation of the remaining unknown field $g^{(2)}_{ij}$ gives rise to
\bea
{\rm Tr} g^{(2)} &=& -\ft{1}{12}\ell^2 (2\hat{R}^{(0)}-\alpha (\nabla^{(0)}\chi^{(0)})^2)\,,
\cr
g^{(2)}_{ij}&=&\ft{1}{12}\ell^2 \big(\hat{R}^{(0)}g^{(0)}_{ij}-6\hat{R}^{(0)}_{ij}\big)\cr
&&
-\ft{1}{24}\ell^2 \Big(\beta_2\,(\nabla^{(0)}\chi^{(0)})^2 \hat{g}^{(0)}_{ij}
 -6(\alpha-\ft{4\gamma}{\ell^2})\nabla^{(0)}_{i}\chi^{(0)}\nabla^{(0)}_{j}\chi^{(0)}\Big)\,.
\eea
Substituting these back into (\ref{actionA}), we find that the on-shell action contains two parts:
\be
\mathcal{A}=\mathcal{A}_{\rm gr}+\mathcal{A}_{m}\,,
\ee
where $\mathcal{A}_{\rm gr}$ is the anomalous term in the gravitational sector, namely
\be
\mathcal{A}_{\rm gr}=\ft{1}{8}\ell^3 (\hat{R}^{(0)}_{ij}\hat{R}^{(0)ij}-\ft{1}{3}\hat{R}^{(0)2})\,.
\label{D5genAgr}
\ee
The quantity $\mathcal{A}_{m}$ denotes the anomalous term associated with the Horndeski scalar, given by
\bea
\mathcal{A}_{m}&&=-\beta_2\Big(\ft{1}{8}\ell^3\big(\hat{R}^{(0)}_{ij}\nabla^{(0)i}\chi^{(0)}\nabla^{(0)j}\chi^{(0)}
-\ft{1}{3}\hat{R}^{(0)}(\partial^{(0)}\chi^{(0)})^2\big)-\ft{1}{16}\ell^3\chi^{(0)}\Box^{(0)2}\chi^{(0)}
\cr
&&-\ft{1}{48}\ell(\ell^2 \alpha-3\gamma)(\nabla^{(0)}\chi^{(0)})^4\Big)
+\ft{1}{2}\hat{R}^{(2)}\,.
\eea
Note that $\hat{R}^{(2)}$ can be obtained explicitly in terms of the scalar field:
\bea
\hat{R}^{(2)}=-\ft{1}{24}\ell^2 (3(\alpha-\ft{2\gamma}{\ell^2})\Box^{(0)}(\nabla^{(0)}\chi^{(0)})^2 -6(\alpha-\ft{4\gamma}{\ell^2})\nabla^{(0)i}\nabla^{(0)j}(\nabla^{(0)}_{i}
\chi^{(0)}\nabla^{(0)}_{j}\chi^{(0)}))\,.
\eea
Thus we see that ${\cal A}_{\rm gr}$ depends only on the boundary curvature invariants,
whilst ${\cal A}_m$ depends only on the boundary scalar invariants.  Intriguingly there is no mixture of the two types, as if $\chi^{(0)}$ is minimally coupled, even though $\chi$ is non-minimally coupled in the bulk. There is a point in parameter space $\alpha=\fft{3\gamma}{\ell^2}$ which suppresses $(\nabla^{(0)}\chi^{(0)})^4$ term.  As we have remarked earlier, $\chi^\0$ is the source of the corresponding scalar operator in the dual field theory. Consequently if we turn off the source, ${\cal A}_m$ vanishes whilst $A_{\rm gr}$ remains unchanged.

The anomaly in the gravitational sector $\mathcal{A}_{\rm gr}$ is associated with the central charges, and it follows from (\ref{D5genAgr}) and (\ref{D5acbb}) that we have
\be
a=c=\dfrac{\ell^{3}}{16}\,.\label{genericac}
\ee
This is precisely the same central charges as in Einstein gravity.  While the Horndeski scalar does give anomalous contribution to ${\cal A}$, it does not change ${\cal A}_{\rm gr}$ in the gravitational sector.

\subsection{Holographic conformal anomaly at critical point}
\label{sec:D=5crit}

The critical point in $D=5$ Einstein-Horndeski gravity is
\be
\gamma=\ft16 \ell^2\alpha\,,
\label{critical5D}
\ee
corresponding to $\beta_2=0$.  As we saw in the previous subsection, the equations become degenerate, indicating that there exist additional modes in the FG expansions. In fact, as we mentioned in the introduction, the theory admits the nearly AdS spacetime, which, in the FG coordinates, is given by
\bea
ds^2 &=& \fft{\ell^2}{4\rho^2}d\rho^2 + \fft{1}{\rho} dx^i dx^i\,,\qquad \chi(\rho) = \chi_s \log\rho + \chi_0\,,\nn\\
\Lambda_0 &=& -\fft{6}{\ell^2}(1+\fft{2\chi_{s}^2 ~\gamma}{\ell^2})\,.\label{lads5}
\eea
Although the metric is exactly the AdS spacetime, the full AdS isometry is broken by the scalar $\chi$.

Thus for asymptotically nearly AdS geometries, the correct FG expansion for the Horndeski scalar contains an additional logarithmic mode, and hence the FG expansion is given by (\ref{criticalchi}).
Note that the logarithmic mode was indeed present in the black hole constructed in \cite{Anabalon:2013oea}. In the notation of \cite{Feng:2015oea}, it is given by $
\left|\chi_{s}\right|=\ft12\ell\sqrt{\beta}$, where $\beta$ is the parameter used in \cite{Feng:2015oea}.

Including the logarithmic mode, we find that for generic couplings the coefficient $\mathcal{A}$ in (\ref{FGaction}) is now given by
\begin{eqnarray}
 \ell^{-1}\mathcal{A}&=& \beta_1 {\rm Tr} g^{(4)}+A_1 {\rm Tr} (g^{(2)2})+A_2({\rm Tr} g^{(2)})^{2}+A_5\big( \hat{R}^{(0)}_{ij}g^{(2)ij}-\hat{R}^{(2)}\big)\nn\\
 &&+{\rm Tr} (g^{(2)})\Big(-\ft12{A_5}\hat{R}^{(0)}+A_3 (\nabla^{(0)}\chi^{(0)})^2
 -\fft{\chi_s \alpha}{\ell^2}\chi^{(2)}\Big)+ A_4 g^{(2)ij}\nabla^{(0)}_i \chi^{(0)}\nabla^{(0)}_j \chi^{(0)}\nn\\
 &&-\fft{4\chi_s \beta_2}{\ell^2} \chi^{(4)}-\fft{\beta_2}{\ell^2} \chi^{(2)2}-\fft{\chi_s \gamma}{\ell^2}\hat{R}^{(0)}\chi^{(2)}
 - \ft{1}{2}\beta_2~ g^{(0)ij}\nabla^{(0)}_i \chi^{(2)}\nabla^{(0)}_j \chi^{(0)} \nn\\
 && +\ft{1}{4}\gamma \hat{G}^{(0)ij}\nabla^{(0)}_i \chi^{(0)}\nabla^{(0)}_j \chi^{(0)}
 \,.
\end{eqnarray}
The coefficients $\beta_2$, $A_3$ and $A_4$ are the same as those given in (\ref{D5coef1}). The remaining
coefficients are modified by $\chi_s$ and given by
\bea
&&\beta_{1}=-\ft{3}{\ell^2}-\ft{\chi_{s}^2}{2\ell^2}(\alpha+\ft{6\gamma}{\ell^2})-\ft12{\Lambda_0}\,,
\qquad A_1=\ft{1}{\ell^2}+\ft{\chi_{s}^2}{4\ell^2}(\alpha+\ft{4\gamma}{\ell^2})+\ft14{\Lambda_0}\,,\cr
&& A_2=-\ft{1}{4\ell^2}-\ft{\chi_{s}^2}{8\ell^2}(\alpha+\ft{2\gamma}{\ell^2})-\ft18{\Lambda_0}\,,\qquad A_5=-\ft{1}{2}(1-\ft{\chi_{s}^2 \gamma}{\ell^2})\,.
\eea
The equation of motion associated with the variation of $\chi^{(4)}$ yields
\be
\beta_2 \chi_s=0\,.
\ee
Thus for generic couplings with $\beta_2\ne0$, we must have $\chi_s=0$, leading to the asymptotic
FG expansion (\ref{normalchi}) for the Horndeski scalar.  At the critical point $\beta_2=0$, on the other hand, the coefficient $\chi_s$ for the logarithmic term does not have to vanish.

The equation of motion associated with the variation of $g^{(4)}_{ij}$ relates the bare cosmological constant with the effective cosmological constant $\Lambda_{\rm eff}=-6/\ell^2$, given by (\ref{lads5}).
The variation of the action with respect to $\chi^{(2)}$ leads to
\be
-\fft{6\chi_s}{\ell^4}({\rm Tr} g^{(2)}+\ft16\ell^2\hat{R}^{(0)})=0\,.
\ee
This equation can be solved by requiring $\chi_s=0$, in which case we obtain the results in the previous subsection for generic $\beta_2$.  Instead we solve this equation by requiring
\be
{\rm Tr} g^{(2)}+\ft16\ell^2\hat{R}^{(0)}=0\,.\label{traceg2}
\ee
It is interesting to note that with this, the equation of motion associated with the variation of $\chi_s$ is automatically satisfied.  Finally, the variation of $g^{(2)}_{ij}$ yields $g_{ij}^{(2)}$. Taking the trace with respect to $g_{ij}^{(0)}$ and making use of (\ref{traceg2}), we can also solve for $\chi^{(2)}$.  The results are given by
\bea
\chi^{(2)}&=&-\fft{\ell^2}{48\chi_s}
\Big(2\chi_{s}^2 \hat{R}^{(0)}+3(\nabla^{(0)}\chi^{(0)})^2\Big)\,,\nn\\
g^{(2)}_{ij} &=&\fft{\ell^2}{12(\ell^2+\chi_{s}^2 \gamma)} \Big((\ell^2-2\chi_{s}^2 \gamma)\hat{R}^{(0)}g^{(0)}_{ij}-6(\ell^2-\chi_{s}^2 \gamma)\hat{R}^{(0)}_{ij}\Big)\nn\\
&&
-\fft{\ell^2 \gamma}{8(\ell^2+\chi_{s}^2 \gamma)} \Big((\nabla^{(0)}\chi^{(0)})^2 g^{(0)}_{ij}
-4\nabla^{(0)}_{i}\chi^{(0)}\nabla^{(0)}_{j}\chi^{(0)}\Big)\,.
\eea
Therefore we have
\bea
\hat{R}^{(2)}&=&\fft{\ell^2 \chi_{s}^2 \gamma}{4(\ell^2+\chi_{s}^2 \gamma)}\Box^{(0)}\hat{R}^{(0)}-\fft{\ell^2 \gamma}{8(\ell^2+\chi_{s}^2 \gamma)} (\Box^{(0)}(\nabla^{(0)}\chi^{(0)})^2\nn\\
&&-4\nabla^{(0)i}\nabla^{(0)j}\nabla^{(0)}_{i}(\chi^{(0)}\nabla^{(0)}_{j}\chi^{(0)}))\,.
\eea
Substituting these back into $\mathcal{A}$, we find again that ${\cal A}$ is the sum of two parts
\be
\mathcal{A}=\mathcal{A}_{\rm gr}+\mathcal{A}_{m}\,,
\ee
where
\bea
\mathcal{A}_{\rm gr}&=&\fft{\ell(\ell^2-3\chi_{s}^2~\gamma)}{8}(\hat{R}^{(0)}_{ij}\hat{R}^{(0)ij}
-\ft{1}{3}\hat{R}^{(0)2})+\fft{\ell^3 \chi_{s}^2 \gamma}{8(\ell^2+\chi_{s}^2 \gamma)}\Box^{(0)}\hat{R}^{(0)}\nn\\
&&+\fft{\ell\chi_{s}^4 \gamma^2 }{2(\ell^2+\chi_{s}^2~\gamma)}(\hat{R}^{(0)}_{ij}\hat{R}^{(0)ij}-\ft{1}{4}(\hat{R}^{(0)2}+\Box^{(0)}
\hat{R}^{(0)}))\,,\nn\\
\mathcal{A}_{m}&=&\fft{\gamma^2 \ell((3(\nabla^{(0)}\chi^{(0)})^2-4\chi_{s}^2 \hat{R}^{(0)})(\nabla^{(0)}\chi^{(0)})^2+16\chi_{s}^2 \hat{R}^{(0)}_{ij}\nabla^{(0)i}\chi^{(0)}
\nabla^{(0)j}\chi^{(0)})}{32(\ell^2+\chi_{s}^2 \gamma)}
\cr && +\fft{1}{16\ell}\fft{\ell^2-\chi_{s}^2 \gamma}{\ell^2+\chi_{s}^2 \gamma}\Big(\Box^{(0)}(\nabla^{(0)}\chi^{(0)})^2-4\nabla^{(0)i}\nabla^{(0)j}
\nabla^{(0)}_{i}(\chi^{(0)}\nabla^{(0)}_{j}\chi^{(0)})\Big)\,.\label{D5Agrm}
\eea
The holographic conformal anomaly for the gravitation sector is quite surprising.  Comparing the results with (\ref{D5acbb}), we find that in addition to the $a$ and $c$ charges, namely
\be
a=c=\fft{\ell^3}{16}(1-\fft{3\chi_{s}^2 \gamma}{\ell^2})\,,\label{D5a=ccrit}
\ee
we have two new charges, which we call $b$-charges.  Specifically, they are given by
\be
b_1=\fft{\ell^3 \chi_{s}^2 \gamma}{8(\ell^2+\chi_{s}^2 \gamma)}\,,\qquad
b_2=\fft{\ell\chi_{s}^4 \gamma^2}{2(\ell^2+\chi_{s}^2 \gamma)}\,.
\label{D5bcrit}
\ee

It is worth noting that the overall coefficient of the term $\Box^{(0)}R^{(0)}$ in ${\cal A}_{\rm gr}$ is
\be
b_1-\ft14 b_2=\fft{\ell\chi_{s}^2 \gamma}{8}\fft{\ell^2-\chi_{s}^2 \gamma}{\ell^2+\chi_{s}^2 \gamma}\,.
\ee
Comparing with the shear viscosity/entropy ratio obtained for the black hole in the Einstein-Horndeski theory \cite{Feng:2015oea}, namely
\be
\fft{\eta}{s}=\fft{1}{4\pi}\sqrt{\fft{\ell^2-\chi_{s}^2 \gamma}{\ell^2+\chi_{s}^2 \gamma}}\,,
\ee
we can thus express the above results in terms of purely anomalous charges:
\be
\fft{\eta}{s}=\fft{1}{4\pi}\sqrt{\fft{2(4b_1-b_2)}{3(\ell^3-16a)}}\,.
\ee
It is worth noticing that the $\eta/s$ combination also appears as a coefficient of the total derivative term of ${\cal A}_m$ in (\ref{D5Agrm}).

As in the previous case, when the source mode $\chi^\0$ vanishes, the anomalous term ${\cal A}_m$ vanishes also, but the quantity ${\cal A}_{\rm gr}$ is completely unchanged. The occurrence of the $b_1$-charge requires further comments.  In a conformal field theory, the trace anomaly of the type $\Box R$ with {\it constant} coefficient can be removed by adding the local counterterm $R^2$ with appropriate constant coefficient.  Intriguingly in the holographic picture, such a term does not even arise in the bulk gravity calculation, when gravity and matter are minimally coupled.  However, in Einstein-Horndeski gravity at the critical point, whose boundary field theory is not conformal, but nearly conformal, the anomalous $\Box R$ term does emerge.  This leads to the question whether this term is removable by adding appropriate counterterms in the field theory.

In the previously studied examples of AdS gravities in literature where matter is minimally coupled, the holographic central charges depend only on the ``fixed'' coupling constants of theory, such as the cosmological constant or the couplings of higher-order curvature invariants.  A distinguishing feature that occurs in Einstein-Horndeski gravity at the critical point is that all the central charges depend also on the integration constant $\chi_s$ of the axionic scalar $\chi$.  It follows that the anomalous $\Box R$ cannot be removed by the local counterterms.  In other words, the scalar hair $\chi_s$ appearing in the $b_1$-charge is not a coupling constant, and it should be viewed as the expectation value of a local field that couples to $\Box R$.

In fact, since the $b_2$-charge is forbidden by any conformal field theory, its emergence also indicates the boundary field theory is not conformal.  It is a reflection of the fact that the background is nearly AdS. In other words, although the metric is AdS, the full AdS symmetry is broken spontaneously by the scalar hair $\chi_s$.  The surviving symmetry is the Poincar\'e group together with the scaling invariance, whilst the special conformal invariance is broken.

\section{Establishing an $a$-theorem in $D=5$}
\label{sec:D=5a}

In the previous section, we studied the holographic conformal anomalies in $D=5$ Einstein-Horndeski gravity. We find that for general couplings, the holographic anomaly in the gravitational sector is identical to that in pure Einstein gravity, given by (\ref{genericac}).  The situation changes significantly at the critical point of the couplings.  In additional to the usual $a=c$ charge, which differs from that in Einstein gravity, new conformal charges $b_1$ and $b_2$ also arise.

In this section, we examine whether there exists an $a$-theorem for each of the above charges. We follow \cite{Myers:2010xs} and consider the cohomogeneity-one domain wall ansatz
\be
ds_{5}^2 = dr^2 + e^{2A(r)} (-dt^2 + dx_{1}^2+dx_{2}^2+dx_{3}^2)\,,\qquad
\chi=\chi(r)\,.\label{domainwall4}
\ee
The AdS vacuum is given by $A(r)=r/\ell$, where $\ell$ is the AdS radius. The function $A(r)$ describe the flow to the AdS ``fixed'' point. For asymptotically AdS domain walls where we adopt the convention that the asymptotic region is located at $r\rightarrow \infty$, the coordinate $r$ is related to the FG coordinate $\rho$ (with flat boundary) by
\be
r=-\ft12{\ell}\log\rho\,.\label{rrhorelation}
\ee

\subsection{No $a$-theorem for generic couplings}
\label{sec:D=5genericnoa}

As we have established in section \ref{sec:D=5generic}, for generic couplings in Einstein-Horndeski gravity, the $a$-charge is given by (\ref{genericac}), depending only on the AdS radius $\ell$.  This leads to a natural choice of the $a$ function
\be
a(r)=\fft{1}{16A'^{3}}\,,
\label{ordinaryaf4}
\ee
which implies that $a'=-3A''/(16A'^4)$.  We now assume that there are additional matter energy-momentum tensor $T_{\mu\nu}^{\rm mat}$ for some generic minimally coupled matter in Einstein-Horndeski gravity, the full Einstein equations of motion becomes
\be
E_{\mu\nu} = T_{\mu\nu}^{\rm mat}\,,
\ee
where $E_{\mu\nu}$ is given by (\ref{EOM}).  Thus for our domain wall ansatz, we have
\bea
-{E}_{t}^t+{E}_{r}^r &=&-3A''-\ft{3}{2}\gamma A'\chi'\chi''+\ft{1}{4}(-2\alpha+12\gamma A'^2-3\gamma A'')\chi'^2\nn\\
&=&-(T^{\rm mat})_{t}{}^t+(T^{\rm mat})_{r}{}^r\,.
\label{null4}
\eea
The null energy condition for the matter field, $-(T^{\rm mat})_{t}{}^t+(T^{\rm mat})_{r}{}^r\ge 0$, implies that
\be
16a' A'^4 (1 + \ft14 \gamma \chi')^2 -\ft32 \gamma A' \chi' \chi'' +
\ft12 (6\gamma A'^2 -\alpha)\chi'^2\ge 0\,.
\ee
Thus we see that for non-vanishing coupling constant $\gamma$ of the Horndeski term, we cannot establish the $a$-theorem $a'\ge 0$. (Note that $a'\ge0$ when $\gamma=0$ for $\alpha\ge 0$, as one would expect.)

\subsection{An $a$-theorem at the critical point}
\label{sec:D=5criticala}

Novel RG flow property arises when coupling parameters are at the critical point. In particular, the equations of motion imply that the Horndeski scalar no longer vanishes in the AdS vacuum (with $\Lambda_{\rm eff}$,) but is given by
\be
\chi(r)=-\fft{2}{\ell}\chi_s r+\chi_0\,.\label{chisol}
\ee
It follows from (\ref{rrhorelation}) that this linear $r$ term corresponds to the logarithmic term in the FG expansion of the Horndeski scalar. As was mentioned earlier, the full AdS isometry of the vacuum is broken down by the scalar to the Poincar\'e group plus the scale invariance.

Since additional matter fields we consider is minimally coupled, the equation of motion for $\chi$ remains $E_\chi=0$ as in (\ref{EOM}). Since in the domain wall ansatz, all fields are functions of $r$, it follows that $E_\chi=0$ implies that
\be
(\alpha g_{rr}-\gamma G_{rr}) \partial^r \chi= (\alpha-6\gamma A'^2)\chi'=C\,,
\ee
where $C$ is an integration constant. Since as $r\rightarrow \infty$, $(\alpha - 6 \gamma A')\rightarrow 0$, it follows that in order for $\chi$ behaves as (\ref{chisol}) for $r\rightarrow \infty$, we must have $C=0$. Thus we have
\be
\alpha-6\gamma A'^2=0\,.\label{chi4}
\ee
This equation keeps the domain wall staying at AdS vacuum $A(r)=r/\ell$ irrespective with the existence of other minimally coupled matters. This seems to suggest that the AdS vacuum remains rigidly ``fixed'' and there can be no RG flow.  This is indeed true for the AdS radius parameter; however, there is another integration constant $\chi_s$ that contributes to the $a$ charge and the scalar $\chi$ can run along the RG flow.  In fact in this critical case, the scalar $\chi$ is not determined by the scalar equation $E_\chi=0$, but rather it is determined by the Einstein equations where non-minimally matter energy momentum tensor contributes.

It is thus natural to replace $\chi_s$ by the function $\chi_s=-\ell\chi'/2$, and substitute this into the $a$-charge (\ref{D5a=ccrit}). We thus propose an $a$-function:
\be
a(r)=\ft1{16}\ell^{3}(1-\ft34\chi'^2 \gamma)\,,\qquad
\Longrightarrow\qquad a'=-\ft{3}{32}\ell^{3}\gamma\chi'\chi''\,.
\label{criticalaf4}
\ee
At the critical point, with the help of (\ref{chi4}), it's easy to see from (\ref{null4}) that
we have
\bea
&&0\le -(T^{\rm mat})_{t}^t+(T^{\rm mat})_{r}^r =-{E}_{t}^t+ {E}_{r}^r=-\fft{3}{2\ell}\gamma\chi'\chi''
=\fft{16}{\ell^4} a'\,.
\label{NEC4}
\eea
Thus the $a$-theorem holds true indeed, namely
\be
a'(r)\geq0\,.
\ee
Hence Einstein-Horndeski gravity satisfies the $a$-theorem at critical point in $D=5$.

As we established in section \ref{sec:D=5crit}, in addition to the $a=c$ charges, Einstein-Horndeski gravity at the critical point has also two $b$-charges.  The physical interpretation of these charges is not clear to us. To understand their property better, we use the same argument and propose two $b(r)$-functions
\be
b_1(r)=\fft{\ell^3 \chi '^2 \gamma}{8 (4+\gamma  \chi '^2)}\,,\qquad
b_2(r)=\fft{\ell^3 \gamma^2 \chi'^4}{8(4+\chi'^2 \gamma)}\,.
\ee
A derivative of these two functions respectively give
\be
b_1'=\fft{\ell^3 \gamma (\chi'^2)'}{2(4+\chi '^2 \gamma)^2}\,,\qquad
b_2'=\fft{\ell^3 \gamma^2 \chi'^2 (8+\chi'^2 \gamma)(\chi'^2)'}{8(4+\chi'^2 \gamma)^2}\,.
\ee
It follows from the null energy condition (\ref{NEC4}) that $b_1$ is monotonously decreasing, namely
\be
b_1' \le 0\,.
\ee
The situation for $b_2'$ is somewhat more complicated. In region $\gamma>0$, it is clear from (\ref{NEC4}) that $(\chi'^2)'\leq0$, therefore $b_2'\leq0$. On the other hand, in region $\gamma<0$, we must have $(\chi'^2)'\geq0$, and hence $\chi'^2 \gamma<0$ is monotonous decreasing. The quantity thus $\chi'^2 \gamma$ reaches its minimum at AdS boundary $r\rightarrow\infty$ in this case, i.e. $\chi'^2 \gamma|_{\infty}=4\gamma\chi_{s}^2/\ell^2$. Thus as long as even this minimum renders $8+\chi'^2 \gamma\geq0$, then $b_2'\geq0$ is true for all the $r$ region. We finally conclude
\bea
&&b_2'\leq0\,,\qquad\hbox{when}\qquad \gamma>0\,,\nn\\
&&b_2'\geq0\,,\qquad\hbox{when}\qquad \gamma<0~~\hbox{and}~~\chi_s^2 \le \fft{2\ell^2}{(-\gamma)}\,.
\eea
It is worth commenting that the critical coupling for $\gamma$ is given by (\ref{critical}) with $d=4$. It is natural to take $\alpha$ for the kinetic term of $\chi$ to be positive, in which case, only the first case above arises.  For completeness, we also give the second case above for which $\alpha$ is negative.

\section{A holographic $a$-theorem in general dimensions}
\label{sec:genD}

In section \ref{sec:D=5a}, we obtain the holographic conformal anomaly for $D=5$ Einstein-Horndeski gravity for general and critical couplings.  In particular, new anomalous terms emerge in the gravitational sector at the critical point.  In general $D=2n+1$ ($d=2n$) dimensions, we expect that the total anomalous terms at the critical point take the form
\be
\mathcal{A}_{\rm gr}=-a E^{(2n)}+\sum_{i} c_{i}I_{i}^{(2n)}+\sum_{k} b_k H_{k}^{(2n)}\,,
\ee
where $H^{(2n)}_k$ are some specific invariants constructed from the boundary curvature tensors and their covariant derivatives. Using the general FG expansion to determine these $(a,b_k,c_i)$ charges is a formidable task.  In fact, we have not obtained the explicit forms for $H^{(2n)}_k$ terms except for $D=5$.  We shall give a conjecture of their defining property presently. In this section, we shall be content to obtain only the $a$ charge, in which case, we can adopt the trick discussed in \cite{Li:2017txk}.  First one note that if the $g_{ij}^{(0)}$ is conformally flat, then all $I_i^{(2n)}$ vanishes.  Furthermore, in $D=5$, there are two $b$-charges, the coefficients of the following two terms
\be
H_{1}^{(4)}=\Box^{(0)}\hat{R}^{(0)}\,,\qquad
H_{2}^{(4)}=\hat{R}^{(0)}_{ij}\hat{R}^{(0)ij}-\ft{1}{4}(\hat{R}^{(0)2}+\Box^{(0)}\hat{R}^{(0)})\,.
\ee
If we choose that $g_{ij}^{(0)}$ to be Einstein, then both $H_1^{(4)}$ and $H_2^{(4)}$ vanish identically.  Thus if we chose a special class of FG expansion where $g_{ij}$ is conformally flat and Einstein, then we can read off the $a$ charge only.  Thus if our task is to read off the $a$ charge only, we can choose a subclass of FG expansion, for which the calculation becomes much simpler.

      In order to obtain the $a$-charge in general dimensions, we make a conjecture that all $H^{(2n)}_k$ vanish when $g_{ij}$ is Einstein. This can be viewed as the defining property of $H_k^{(2n)}$.

\subsection{The $a$-charge in higher dimensions}
\label{genDa}

The simplest Einstein metric that is conformally flat is that for a round sphere. We follow \cite{Li:2017txk} and consider the reduced FG expansion, with the metric
\be
ds^2=\fft{\ell^2}{4\rho^2}d\rho^2+f(\rho)d\Omega_{d}^2\,,
\ee
where $d\Omega_d^2$ is the metric for the round $d$-sphere $S^d$. The function $f(\rho)$ and the Horndeski scalar $\chi(\rho)$ expand as
\bea
f &=& f_0 +f_2 \rho+f_4 \rho^2+f_6 \rho^3+\cdots\,,\nn\\
\chi&=&\chi_s \log\rho+\chi_2 \rho+\chi_4 \rho^2+\chi_6 \rho^3+\cdots\,,
\eea
where $f_{2i}$, $\chi_{2i}$ and $\chi_s$ are all constant variables. Now in $D=2k+1$, the anomaly ${\cal A}$ is a functional of $(f_2 ,f_4 ,\cdots, f_{2k})$,  $(\chi_2 ,\chi_4 ,\cdots, \chi_{2k})$ and $\chi_s$. In particular the terms involving $\chi_{2k}$ and $f_{2k}$ are linear in these two variables, given by
\bea
{\cal A}&=&-\fft{d}{2\ell^3 f_0}
\Big(\ell^4 \Lambda_0+\fft{d(d-1)}{2}\gamma\chi_{s}^2 +\ell^2 (\ft12d(d-1)+\alpha\chi_{s}^2)\Big)f_{2k}
\cr
&&-\fft{d}{\ell^3}\big(\ell^2 \alpha-\ft12d(d-1)\gamma\big)\chi_s \chi_{2k} + \cdots\,,
\eea
The variation of $\chi_{2k}$ implies
\be
\Big(\alpha \ell^2- \ft12 d(d-1)\gamma\Big)\, \chi_s=0\,.\label{genDchoice}
\ee
The solution is either $\chi_s=0$ or the taking the critical condition (\ref{critical}).  The former leads to an $a$-charge that is identical to that in Einstein gravity and the Horndeski term gives no contribution.  As in the $D=5$ case discussed in section \ref{sec:D=5a}, we can find no $a$-function for non-vanishing Horndeski coupling that satisfies the $a$-theorem.

    The situation becomes more interesting if we solve (\ref{genDchoice}) by requiring the critical condition (\ref{critical}).  At the critical point, the variation of $f_{2k}$ leads to
\be
\Lambda_0=-\fft{d(d-1)}{2\ell^2}(1+\fft{2\chi_{s}^2 \gamma}{\ell^2}).
\ee
Then the variation of the remainder variables $(f_2 ,f_4 ,\cdots, f_{2(k-1)})$ and $(\chi_2 ,\chi_4 ,\cdots, \chi_{2(k-1)})$ implies that
\bea
&& \chi_{2n}=(-1)^n \fft{\ell^{2n}\chi_s}{2^{2n-1}nf_{0}^{n}}\,,\qquad 1 \le n\le k-1\,.
\cr &&
\cr && f_2=-\ft12\ell^2\,,\qquad f_4=\fft{\ell^4}{16 f_0}\,,\qquad f_{2n}=0\,,\qquad 3 \le n\le k-1\,.
\eea
We can then read off the $a$-charge, up to an overall numerical constant, as
\be
a=\ell^{d-1}\Big(1-(d-1)\fft{\chi_{s}^2\gamma}{\ell^2}\Big)\,.
\label{centralcharge}
\ee
In addition to the round sphere boundary, we verified that many other boundaries can also give (\ref{centralcharge}), as long as they are Einstein metrics. This is because when bulk theory involves only Ricci tensors, all the $c$-charges are proportional to the $a$-charge, giving rise to a net $a$-charge, up to pure numerical factor \cite{Li:2017txk}.  The same outcome of the $a$-charge for various different Einstein metrics as boundaries confirms our original conjecture that the $b$-charge terms all vanish when the boundary metrics are Einstein.

\subsection{An $a$-theorem in general dimensions}
\label{sec:genDat}

As discussed in section \ref{sec:D=5a} for $D=5$, we consider here the dimain wall ansatz in general dimensions:
\be
ds_{d+1}^2 = dr^2 + e^{2A(r)} (-dt^2 + dx_i dx^i)\,,\qquad\chi=\chi(r)\,.\label{domainwall}
\ee
At the critical point, For the same argument in $D=5$, we find that the equation of motion for the Horndeski scalar is reduced to
\be
\alpha g_{rr}-\gamma G_{rr}=\alpha-\ft12{d(d-1)}\gamma A'^2=0\,.
\label{chi}
\ee
Thus the AdS geometry is fixed and not running.  We thus instead substitute $\chi_s =-\ell\chi'/2$ into the general $a$-charge (\ref{centralcharge}) and define an $a$-function at critical point
\be
a(r)=\ell^{d-1}\big(1-\ft14{(d-1)\chi'^2 \gamma}\big)\,,
\label{criticalaf}
\ee
which implies that
\be
a'(r)=-\ft12(d-1)\ell^{d-1}\gamma\chi'\chi''\,.
\ee
The null energy condition in general dimensions implies that
\bea
&&-(T^{\rm mat})_{t}{}^t+(T^{\rm mat})_{r}{}^r =-{E}_{t}^t+{E}_{r}^r =-(d-1)A''-\ft12(d-1)\gamma A'\chi'\chi''\cr &&+\ft{1}{4}\Big(-2\alpha+d(d-1)\gamma A'^2-(d-1)\gamma A''\Big)\chi'^2 \geq0
\label{null}
\eea
At the critical point, using (\ref{chi}), (\ref{null}) becomes
\bea
&&-(T^{\rm mat})_{t}{}^t+(T^{\rm mat})_{r}{}^r =-\mathcal{E}_{t}^t+\mathcal{E}_{r}^r=-\fft{d-1}{2\ell}\gamma\chi'\chi''
=\fft{a'}{\ell^d}\geq0\,.
\eea
Thus the $a$-theorem
\be
a'(r)\geq0
\ee
is established at the critical point with the null energy condition.

\subsection{Holographic conformal anomaly in $D=3$}
\label{sec:D=3}

In section \ref{sec:D=5}, we considered the $D=5$ example and used the full FG expansion to construct the holographic conformal anomalies. For general dimensions, the full FG expansion becomes very complicated and we adopted the technique of \cite{Li:2017txk} and used reduced FG expansion ansatz for $D\ge 7$ dimensions.

For $D=3$, the situation becomes simpler than that in $D=5$ and we can easily analyze the holographic conformal anomaly using the full FG expansions (\ref{FG}) and (\ref{criticalchi}). We find
\be
S=\fft{1}{16\pi}\int d^{2}x\int_{\epsilon} d\rho\sqrt{-g}\mathcal{L}=\fft{1}{16\pi}\int d^{2}x \sqrt{g^{(0)}}\int_{\epsilon} d\rho(\cdots+\fft{\mathcal{A}}{\rho}+\cdots)\,,
\ee
where the quantity $\mathcal{A}$ is
\be
\ell^{-1}\mathcal{A}=A_1 \hat{R}^{(0)}+A_2 {\rm Tr}g^{(2)}+A_3 \chi^{(2)}+A_4 (\partial^{(0)}\chi^{(0)})^2\,.
\ee
Various coefficients $A_i$ above are
\bea
A_1 &=& \fft{\ell^2-\chi_{s}^2 \gamma}{2\ell^2}\,,\qquad
A_2=-\fft{\ell^2+\chi_{s}^2 (\gamma+\ell^2 \alpha)+\ell^4 \Lambda_0}{2\ell^4}\,,\nn\\
A_3 &=& \fft{2\chi_{s}(\gamma-\alpha\ell^2)}{\ell^4}\,,\qquad
A_4=\fft{\gamma-\alpha\ell^2}{4\ell^2}\,.
\eea
The variations of $\chi^{(2)}$ and $g^{(2)}_{ij}$ provide the critical condition and determine the cosmological constant:
\be
\alpha=\fft{\gamma}{\ell^2}\,,\qquad
\Lambda_0=-\fft{1}{\ell^2}(1+\fft{2\chi_{s}^2 \gamma}{\ell^2})\,.
\ee
Then the holographic conformal anomaly in $D=3$ is simply
\be
\mathcal{A}=\ft12{\ell}(1-\fft{\chi_{s}^2 \gamma}{\ell^2})\hat{R}^{(0)}\,.
\ee
Interestingly, there is no anomaly in the matter sector at all at the critical point. The $a$-charge is given by
\be
a=\ell(1-\fft{\chi_{s}^2 \gamma}{\ell^2})\,.
\ee
This gives (\ref{centralcharge}) with $d=2$.  The corresponding $a$-theorem was established in the previous subsection.

\section{Conclusion}

In this paper, we studied Einstein-Horndeski gravity with the Horndeski terms involving only up to and including the linear curvature tensor.  We took the cosmological constant to be negative so that the theory admitted the AdS spacetimes.  The theory may be expected to be dual to some strongly coupled conformal field theory in the AdS boundary, by the virtual of the AdS/CFT correspondence.  Applications in the AdS/CMT correspondence were studied in literature.  In this paper, we analyzed the holographic conformal anomaly and studied the corresponding $a$-theorem.

We focused our discussion on $D=5$ bulk dimensions, corresponding to $d=4$ boundary dimensions. We adopted the full complete FG expansion to derive the conformal anomalies.  We find that the total anomaly splits into two parts. One is the anomaly in gravitational sector, given by the curvature tensor
invariants of the boundary metric.  The other is the anomaly in the matter sector, given by the invariants constructed from the Horndeski scalar.  There is no mix between the two sectors as if the theory were minimally coupled.  For the generic Horndeski coupling, we found that the anomaly in the gravitational sector was the same as that in Einstein gravity, and the Horndeski scalar gave no contribution to the $a=c$ charges.  There exists a critical point of the Horndeski coupling for which case the theory admits a nearly AdS spacetime with the non-vanishing scalar.  The full AdS isometry is broken by the scalar hair to the Poincar\'e group plus the scale invariance. The corresponding scale-invariant quantum field theory has a much richer anomalous structure.  Note only the Horndeski scalar contributes to the $a=c$ charges, it generates two additional $b$-charges in the gravitational sector.

Intriguingly, the boundary curvature polynomials $H^{(4)}$ associated with the $b$-charges vanish identically for any Einstein metrics. Although we have not computed the $H^{(2n)}$ explicitly except for $D=5$, we conjectured that this would be true in general $D=2n+1$ dimensions.  In fact this may be the defining property for $H^{(2n)}$.  This conjecture allowed us to consider general odd dimensions and use reduce FG expansions to deduce the $a$-charge by taking the boundary metrics in the FG expansion to be Einstein.  We obtained the general formula for the $a$-charge at the critical point in general odd bulk dimensions. We used a variety of different Einstein metrics and obtained same result for the $a$-charge, confirming our conjecture.

We then used the domain wall ansatz to study the holographic $a$-theorem.  For generic Horndeski couplings, we failed to find an $a$-function that would satisfy the $a$-theorem.  On the other hand, for the critical couplings, we found that the $a$-theorem could be established for general odd bulk dimensions.  In this case, the $a$-charge depends on two parameters, one is the AdS radius, and the other is an integration constant of the Horndeski scalar.  We found that at the critical point, the AdS radius was rigidly fixed and the domain wall was the rigidly AdS metric, but the Horndeski scalar could run with the holographic RG flow.  In $D=5$, we also obtained explicit expressions for the two new $b$-charges, and we found that there existed monotonous $b$-functions.  The physical meanings of these two $b$-charges, and their apparent ``$b$-theorems'' remain to be further investigated.

Not all gravity theories yield sensible conformal or nearly conformal field theory under the AdS/CFT correspondence. The dual conformal field theory of Einstein-Horndeski gravity is far from clear.  Our results provide encouraging signs for the theory at the critical point.

\section*{Acknolwedgement}

We are grateful to the referee for raising questions on the holographic $b_1$-charge, and to Zhao-Long Wang for useful discussions to clarify the issues. This work is supported in part by NSFC grants No.~11475024 and No.~11235003.

\end{document}